\documentclass[11pt]{article}
\usepackage[final]{acl}

\usepackage{times}
\usepackage{latexsym}

\usepackage[T1]{fontenc}
\usepackage[utf8]{inputenc}

\usepackage{microtype}

\usepackage{inconsolata}

\usepackage{graphicx}

\usepackage{amsmath,amssymb} 
\usepackage{booktabs}
\usepackage{dsfont}
\usepackage{multirow}
\usepackage[most]{tcolorbox}
\usepackage{verbatim}
\usepackage[table]{xcolor}
\usepackage{xspace}
\usepackage[linesnumbered,ruled,vlined]{algorithm2e}

\newtcolorbox{promptbox}[2][]{
  colback=blue!5!white,
  colframe=blue!75!black,
  fonttitle=\bfseries,
  title=#2,
  breakable,
  #1
}

\newcommand{\methodname}{BootTrans\xspace}

\title{Bootstrapping Code Translation with Weighted Multilanguage Exploration}

\author{
    Yuhan Wu$^\dagger$ \quad 
    Huan Zhang$^\dagger$ \quad 
    Wei Cheng$^\dagger$ \quad 
    Chen Shen$^\dagger$ \quad 
    Jingyue Yang$^\dagger$ \quad 
    Wei Hu$^{\dagger,\,\ddagger,\,}$\thanks{\,\, Corresponding author} \\
    $^\dagger$ State Key Laboratory for Novel Software Technology, Nanjing University, China \\
    $^\ddagger$ National Institute of Healthcare Data Science, Nanjing University, China \\
    \texttt{\{yhwu,zhanghuan,wchengcs,cshen,jyyang\}.nju@gmail.com, whu@nju.edu.cn} 
}

\begin{document}
\maketitle
\begin{abstract}
Code translation across multiple programming languages is essential yet challenging due to two vital obstacles: scarcity of parallel data paired with executable test oracles, and optimization imbalance when handling diverse language pairs. 
We propose \methodname, a bootstrapping method that resolves both obstacles. Its key idea is to leverage the functional invariance and cross-lingual portability of test suites, adapting abundant pivot-language unit tests to serve as universal verification oracles for multilingual reinforcement learning (RL) training. Our method introduces a dual-pool architecture with seed and exploration pools to progressively expand training data via execution-guided experience collection.
Furthermore, we design a language-aware weighting mechanism that dynamically prioritizes harder translation directions based on relative performance across sibling languages, mitigating optimization imbalance. 
Extensive experiments on the HumanEval-X and TransCoder-Test benchmarks demonstrate substantial improvements over baseline LLMs across all translation directions, with ablation studies validating the effectiveness of both bootstrapping and weighting components.
\end{abstract}

%====================%
\section{Introduction}

Large Language Models (LLMs) have shown remarkable progress in coding tasks, revolutionizing contemporary software engineering workflows.
Code translation, migrating code from a source programming language to a target while ensuring the syntax and semantics correctness, is pivotal for legacy system modernization and cross-platform interoperability~\cite{Nguyen2014Migrating,roziere2020transcoder}.
Despite the advancements, code translation usually relies on abundant parallel code of high quality, which may not always be available. 
Even when available, they are rarely equipped with aligned, executable test cases ~\cite{roziere2022leveraging,jiao2023evaluation,Zhu2024semi}. 

To resolve such reliance, existing works \cite{huang2023program, liu2023syntax, szafraniec2022code} explore code structure information to learn representations for unsupervised translation. 
However, they typically demand enormous amounts of monolingual corpora to establish robust cross-lingual alignment. 
Moreover, these methods generally do not leverage executable test cases during training, and thus cannot directly optimize translation quality based on functional correctness. 
As a result, the learning objective is often restricted to syntactic conversion rather than functional equivalence.

Recently, Reinforcement Learning from Verifiable Rewards (RLVR) offers a promising paradigm shift by optimizing models based on execution feedback ~\cite{le2022coderl,shojaee2023executionbased,jana2024cotran}. 
While high-quality parallel code is scarce, unit tests are inherently transferable~\cite{roziere2022leveraging}. 
As unit tests often follow template-based patterns, they allow for highly reliable translation through rule-based methods \cite{cassano2022multipl-e}, and thus provide a viable way to guarantee consistent functional verification across different programming languages.
This observation suggests that by translating test oracles from a resource-rich language (e.g., Python) to target languages, we can construct a rigorous RL environment for multilingual translation without ground-truth references.

\begin{figure}
    \centering
    \includegraphics[width=\linewidth]{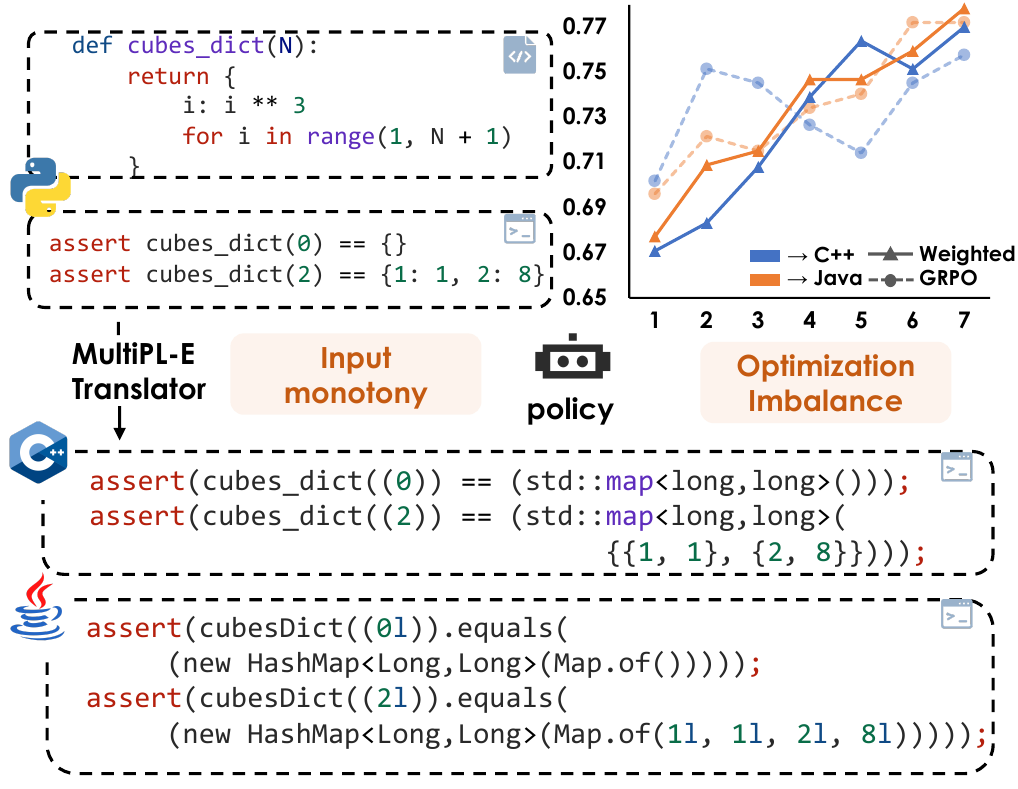}
    \caption{Illustration of challenges in scaling RLVR for multilingual code translation. (i) Input monotony: verifiable seeds are predominantly confined to a single pivot language. (ii) Optimization imbalance: varying task difficulties lead to biased learning signals.}
    \label{fig:challenge}
\end{figure}

However, realizing this potential in multilingual code translation presents two fundamental challenges, as illustrated in Figure~\ref{fig:challenge}. 
First, there is a severe shortage of multilingual datasets that provide unit tests across diverse programming languages to serve as starting points for RLVR.
While test oracles can be migrated from a resource-rich language, high-quality code paired with these oracles remains predominantly available in a single pivot language. 
It is rare to find verified, functionally equivalent source code in other target languages, so only unidirectional translation is enabled.
Relying solely on a static source language dataset leaves a critical void in training data for reverse direction translations. 
While existing benchmarks like HumanEval-X~\cite{zheng2023codegeex} provide multilingual source code, their scale remains limited for training robust models.
Although synthetic data offer a viable workaround, they are intrinsically constrained by reduced diversity and potential bias. 
In addition, scaling up high-quality code across various programming languages remains computationally expensive and technically demanding.
Bootstrapping verified training source code in all target languages is a key for adapting RL to code translation~\cite{yan2023codetransocean,wang2024repotransbench}.

Second, optimizing for multiple programming languages simultaneously introduces optimization imbalance. 
Different translation directions (e.g., Python$\to$Java vs. Python$\to$C++) exhibit varying levels of difficulty due to syntactic and semantic discrepancies~\cite{zhu2022multilingual,yan2023codetransocean,du2024joint}. 
When optimizing uniformly across these tasks, the model tends to be dominated by easier translation directions where rewards are more readily accessible. 
Thus, the model rapidly improves on easier language pairs but often exhibits performance oscillation or stagnation on harder ones, causing suboptimal multilingual proficiency.

In this work, we address these challenges by proposing \methodname. 
To overcome data sparsity, we leverage one language (e.g., Python) as a strategic pivot with abundant source code accompanied by unit tests and propagate test oracles to other languages. 
Then, we expand the RL curriculum through experience collection by utilizing verified rollouts from the policy model.
% This presents a unique opportunity to propagate verification oracles to other languages. 
To mitigate optimization imbalance, we introduce a language-aware weighting optimization mechanism that dynamically adjusts the learning focus based on the relative difficulty and performance of each target language. 
We conduct extensive experiments on pairwise code translation among C++, Java, and Python.
Results demonstrate that \methodname outperforms existing open-source LLMs.

Our main contributions are outlined as follows:
\begin{itemize}
    \item We leverage a resource-rich pivot language to bootstrap a verifiable multilingual corpus, effectively overcoming the reliance on parallel code in code translation.
    
    \item We design a language-aware weighting optimization mechanism to mitigate optimization imbalance across translation directions by dynamically adjusting learning focus for different target languages.
    
    \item We conduct extensive experiments on pairwise code translation among C++, Java, and Python, showing that \methodname consistently outperforms its corresponding base model by up to 26.82\% on the HumanEval-X dataset and 7.46\% on the TransCoder-Test dataset. Code is accessible at \url{https://github.com/nju-websoft/BootTrans/}.
\end{itemize}

\begin{figure*}
    \centering
    \includegraphics[width=\linewidth]{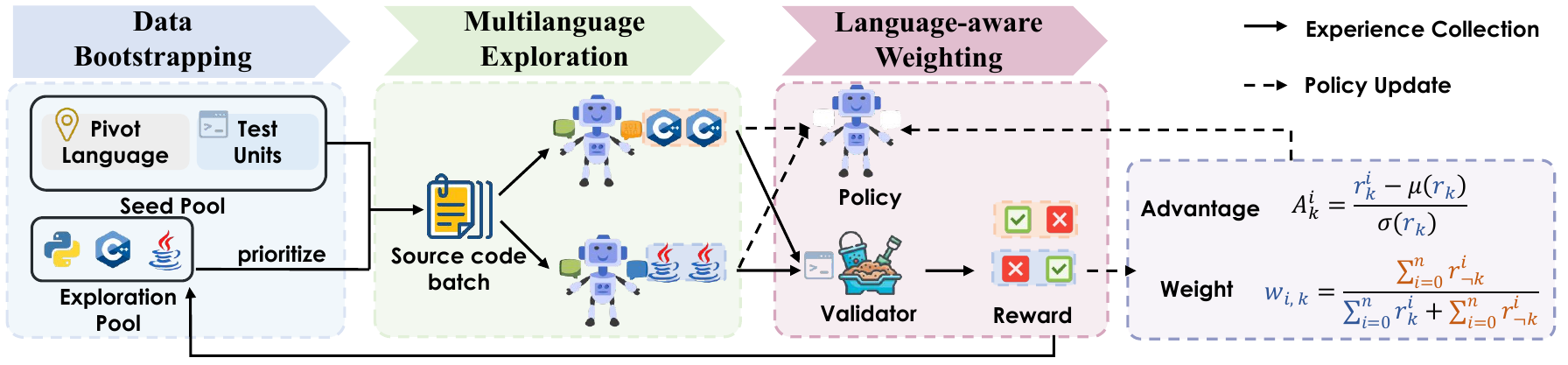}
    \caption{Overview of \methodname. It comprises two key components: (i) \textbf{Bootstrapping Multilanguage Exploration} (left and center panels), which expands the training set via execution-verified translations; (ii) \textbf{Language-aware Weight Optimization} (right panels), which dynamically re-weights the loss via cross-lingual performance gaps.}
    \label{fig:method}
\end{figure*}

%====================%
\section{Problem Formulation}

We study the problem of multilingual code translation across a set of programming languages $\mathcal{L} = \{L_1, \dots, L_M\}$. 
The goal is to learn a unified policy $\pi_\theta(y \,|\, x, L_\text{src}, L_\text{tgt})$ translating a source code $x$ in language $L_\text{src} \in \mathcal{L}$ to a functionally equivalent target code $y$ in language $L_\text{tgt} \in \mathcal{L}$ ($L_\text{src} \neq L_\text{tgt}$).

Unlike traditional supervised settings that rely on parallel corpora for language pairs, we operate under a \textit{monolingual-pivot bootstrapping} scenario. 
We assume access to a seed dataset $\mathcal{D}_{\text{seed}}$ consisting of code-test pairs $\{(x, T)\}$ solely in a pivot language $L_\text{pivot} \in \mathcal{L}$ (e.g., Python). 
The set of unit tests $T$ serves as a transferable verification oracle.
Our objective is to leverage $\mathcal{D}_{\text{seed}}$ to progressively explore the multilingual space and optimize $\pi_\theta$ to maximize the expected success rate of translation across all directions in $\mathcal{L} \times \mathcal{L}$. 

%====================%
\section{Our Method}
Figure~\ref{fig:method} shows the proposed method consisting of bootstrapping multilanguage exploration and language-aware weight optimization.

\subsection{Bootstrapping Multilanguage Exploration}
To break the dependency on curated parallel corpora, we formulate the RL training for code translation as an evolving exploration process.
Instead of confining the model to static pivot code-test pairs, the bootstrapping mechanism empowers the model to curate its own rollouts through execution-guided experience collection. 
These verified experiences serve as augmented source inputs, enabling the model to unlock training for translation directions originating from non-pivot languages.

\subsubsection{Dual-Pool Architecture}
We maintain two distinct data pools to manage the training curriculum:

\textbf{Seed pool} ($\mathcal{D}_{\text{seed}}$).
This pool contains the original dataset $\mathcal{D}_{\text{seed}}= \{(x, T) \,|\, x \in L_\text{pivot} \}$, where $x$ is the source code in the pivot language and $T$ is the corresponding accessible test suite for all languages. 
In fact, $T$ is readily obtainable as they only need to remain consistent with the execution behavior of $x$. Diverse and edge-case-rich inputs can be synthesized using automated techniques such as fuzzing or LLM-based synthesis. 
The corresponding canonical outputs are then obtained by executing $x$ on these inputs, serving as the function output.
Moreover, such a test suite is highly portable to other programming languages by rule-based conversion \cite{cassano2022multipl-e}, which naturally establishes a unified execution environment for RL.
For example, we leverage Python's rich resource as a seed, providing a cold start for exploring the solution space of target languages.

\textbf{Exploration pool} ($\mathcal{D}_{\text{explore}}$).
This pool dynamically stores the policy model's successful translations during rollout.
Formally, we define $\mathcal{D}_{\text{explore}} = \{ (y, T) \,|\, y \in \mathcal{L} \setminus \{L_\text{pivot}\} \}$, where $y$ is a generated code that has passed all test cases in $T$. 
To ensure efficient exploration and prevent pool saturation, we manage $\mathcal{D}_{\text{explore}}$ as a First-In-First-Out (FIFO) queue with a capacity of $|\mathcal{L} \setminus \{L_{pivot}\}|$ times the rollout batch size. 
To prevent semantic drift, only verified rollouts stemming from $D_\text{seed}$ are enqueued, ensuring execution consistency between the test suites and the evolving data pool.

In each training step, we prioritize drawing the full batch from $\mathcal{D}_{\text{explore}}$.
If the remaining items are insufficient, the batch is supplemented with examples from the seed pool $\mathcal{D}_{\text{seed}}$. 
This strategy compels the policy model to exhaustively exploit the current exploration frontier for a given pivot language code before introducing new code-test pairs in $\mathcal{D}_{\text{seed}}$.

\subsubsection{Verification Oracle and Reward}
We optimize the policy model using execution feedback from unit tests. Given a generated candidate $y$ and its associated test suite $T$ in the target language, we define a binary verifiable reward:
\begin{equation}
\resizebox{.89\columnwidth}{!}{$
R(y, T) = \mathds{1}\Big[y \text{ compiles and passes all tests in } T\Big],
\label{eq:reward}
$}
\end{equation}
where $\mathds{1}$ denotes the indicator function.
In practice, compilation errors, runtime errors, and timeouts yield $R=0$. This setting aligns the optimization objective with functional correctness rather than surface-form similarity.

\subsubsection{Multilingual Expansion Training}
Algorithm~\ref{alg:boostrans} shows the training process.
We structure the exploration as a progressive expansion rooted in the pivot language, which follows a standard RL pipeline, consisting of rollout, reward computation, and policy update. 
In each training iteration, we construct a batch by sampling source code $x$ from both $\mathcal{D}_{\text{seed}}$ and $\mathcal{D}_{\text{explore}}$.
Given $x$ in language $L_\text{src}$, we obtain rollouts from the policy $\pi_\theta$ by generating $G$ candidate translations $\mathcal{Y}_k = \{y_{k}^1,$ $\dots, y_{k}^G\}$ for each target language $L_k \in \mathcal{L} \setminus \{L_\text{src}\}$.
We verify these candidates against the test oracle $T$ and compute rewards for $\mathcal{Y}_k$.
Subsequently, we perform experience collection to update the exploration pool $\mathcal{D}_{\text{explore}}$ immediately following the reward computation.
If multiple candidates in $\mathcal{Y}_k$ pass $T$, we randomly retain one for $\mathcal{D}_{\text{explore}}$. Conversely, if no candidate passes $T$, no entry is added, pruning the exploration of tasks beyond the model’s current reach.
These verified translations are eligible to be sampled as \textit{source} inputs in subsequent iterations.
This enables the model to learn reverse translations (e.g., Java$\to$Python) and cross-lingual translations (e.g., Java$\to$C++) that are absent in the seed data.
As a result, we progressively populate the $\mathcal{L} \times \mathcal{L}$ translation task matrix.

\begin{algorithm}[!t]
\caption{\methodname}
\label{alg:boostrans}
\SetKwComment{Comment}{$\triangleright$\ }{}
\newcommand\mycommfont[1]{\footnotesize\ttfamily\textcolor{blue}{#1}}
\SetCommentSty{mycommfont}
\SetNoFillComment
\small
\KwIn{Seed dataset $\mathcal{D}_\text{seed}$, test units $T$, initial policy $\pi_{\theta_\text{init}}$, hyperparameters $N,B,G$}
\KwOut{Optimized policy $\pi_\theta$}
        $\pi_\theta \leftarrow \pi_{\theta_\text{init}}, \mathcal{D}_\text{explore} \leftarrow \emptyset$\;
        \For{$step \leftarrow 1$ \KwTo $N$}{
            \Comment{Sample a batch of source code}
            $X \leftarrow \{x_1, \dots, x_B\}$ from $\mathcal{D}_\text{seed} \cup \mathcal{D}_\text{explore}$\; 
            \ForEach{source code $x_i \in X$}{
                Let $L_\text{src}^i$ be the language of $x_i$, define target languages $\mathcal{L}_\text{tgt}^i = \mathcal{L} \setminus \{L_\text{src}^i\}$\;
                \ForEach{target language $L_k \in \mathcal{L}_\text{tgt}^i$}{
                    \Comment{Generate $G$ candidates}
                    $\{y_{i,k}^1, \dots, y_{i,k}^G\} \sim \pi_{\theta}(\cdot \,|\, x_i, L_k)$\;
                    \Comment{Verify and reward computation}
                    Get $R(y_{i,k}^j, T)$ using Eq.~\eqref{eq:reward}\;
                    \Comment{Expand exploration pool}
                    $\mathcal{D}_\text{explore} \leftarrow \cdot \cup \{y_{i,k}^j \,|\, R(y_{i,k}^j, T) = 1\}$; 
                }
                \ForEach{$L_k \in \mathcal{L}_\text{tgt}^i$}{
                    \Comment{Compute cumulative reward}
                    $\mathcal{R}_{i,k} \leftarrow \sum_{j=1}^G R(y_{i,k}^j, T)$\;
                    \Comment{Compute sibling reward}
                    $\mathcal{R}_{i,\neg k} \leftarrow \sum_{L_j \in \mathcal{L}_\text{tgt}^i, j \neq k} \mathcal{R}_{i,m}$\;
                    \Comment{Compute language-aware weight}
                    $w_{i,k} \leftarrow \frac{\mathcal{R}_{i,\neg k}}{\mathcal{R}_{i,k} + \mathcal{R}_{i,\neg k}}$\; 
                }
            }
            Update $\pi_\theta$ by maximizing Eq.~\eqref{eq:grpo-loss}\;
        }
\end{algorithm}

\begin{table*}
    \centering
    \resizebox{\textwidth}{!}{
    \begin{tabular}{l|cccccc|c|cccccc|c}
        \toprule
        \multirow{2}{*}{LLMs} & \multicolumn{7}{c|}{HumanEval-X} & \multicolumn{7}{c}{TransCoder-Test} \\ 
        \cmidrule(lr){2-8} \cmidrule(lr){9-15} & P$\to$J & P$\to$C & J$\to$P & J$\to$C & C$\to$J & C$\to$P & Avg & P$\to$J & P$\to$C & J$\to$P & J$\to$C & C$\to$J & C$\to$P & Avg \\
        \midrule
        Qwen3-1.7B & 54.27 & 43.29 & 82.32 & 58.54 & 75.00 & 72.56 & 64.33 & 75.52 & 76.66 & 80.17 & 81.37 & 83.61 & 80.17 & 79.58 \\
        \rowcolor{gray!15} Qwen3-32B & 68.29 & \textbf{64.63} & 86.59 & 62.20 & 60.98 & 65.24 & 67.99 & 57.47 & 67.88 & 78.02 & 72.81 & 59.75 & 76.08 & 68.67 \\
        \methodname Qwen3-1.7B & \textbf{73.78} & 60.37 & \textbf{87.20} & \textbf{70.73} & \textbf{77.44} & \textbf{78.66} & \textbf{74.70} & \textbf{79.88} & \textbf{85.87} & \textbf{83.62} & \textbf{91.86} & \textbf{84.85} & \textbf{82.11} & \textbf{84.70} \\
        
        \midrule
        Llama-3.1-8B-Instruct & 37.20 & 57.32 & 84.76 & 47.56 & 73.17 & 70.73 & 61.79 & 75.10 & 79.44 & 75.00 & 87.58 & 80.29 & 72.20 & 78.27 \\
        \rowcolor{gray!15} Llama-3.1-70B-Instruct & \textbf{87.80} & \textbf{78.66} & \textbf{86.59} & \textbf{83.54} & \textbf{87.20} & \textbf{84.76} & \textbf{84.76} & 79.25 & \textbf{90.15} & 80.39 & 90.15 & 81.95 & 79.53 & 83.57 \\
        \methodname Llama-3.1-8B-Instruct & 73.78 & 66.46 & 85.98 & 76.83 & 84.76 & 82.32 & 78.36 & \textbf{78.01} & 86.51 & \textbf{84.05} & \textbf{92.29} & \textbf{82.78} & \textbf{81.03} & \textbf{84.11} \\
        
        \midrule
        Qwen2.5-7B-Instruct & 51.22 & 69.51 & 86.59 & 59.15 & 64.02 & 80.49 & 68.50 & 86.51 & 88.87 & 87.72 & 92.29 & 89.00 & 84.91 & 88.22 \\
        \rowcolor{gray!15} Qwen2.5-32B-Instruct & 62.80 & 74.39 & \textbf{90.85} & 55.49 & 65.24 & 82.32 & 71.85 & 86.72 & \textbf{92.51} & \textbf{89.22} & \textbf{93.15} & 89.00 & 85.99 & \textbf{89.43} \\
        \methodname Qwen2.5-7B-Instruct & \textbf{81.71} & \textbf{76.83} & 90.24 & \textbf{82.32} & \textbf{89.02} & \textbf{82.93} & \textbf{83.84} & \textbf{86.72} & 89.72 & 88.79 & 92.72 & \textbf{90.87} & \textbf{86.64} & 89.24 \\
        \bottomrule
    \end{tabular}}
    \caption{CA@1 scores on HumanEval-X and TransCoder-Test benchmarks. 
    ``C'', ``J'', and ``P'' denote C++, Java, and Python, respectively.
    For each model family, we show: (i) base model, \colorbox{gray!15}{(ii) large-scale reference (shaded),} and (iii) our method \methodname (\textbf{bold when highest}).
    ``Avg'' denotes the average score across six translation directions.}
    \label{tab:model_main}
\end{table*}

\subsection{Language-aware Weight Optimization}
To address the imbalance in optimization caused by varying translation capabilities across target languages, we introduce a language-aware weight optimization mechanism. 
Intuitively, the learning signal for a specific target language $L_k$ should be amplified when the model underperforms on $L_k$ despite demonstrating high proficiency in other ``sibling'' languages for the same source code. 
% Formally, for each source code $x_i$ in language $L_\text{src}$, the policy generates translations for a set of target languages $\mathcal{L}_\text{tgt}^i = \mathcal{L} \setminus \{L_\text{src}\}$.

For each specific target language $L_k \in \mathcal{L}_\text{tgt}^i$ for source code $x_i$, let $\mathcal{R}_{i,k} = \sum_{y \in \mathcal{Y}_{i,k}} R(y, T)$ denote the cumulative reward obtained by the candidate group $\mathcal{Y}_{i,k}$.
We define the sibling reward $\mathcal{R}_{i,\neg k} = \sum_{L_j \in \mathcal{L}_\text{tgt}^i, j \neq k} \mathcal{R}_{i,j}$ as the aggregate performance on all other target languages.
The optimization weight $w_{i,k}$ for the translation task $L_\text{src} \to L_k$ is computed as:
% $w_k = \frac{\mathcal{R}_{\neg k}}{\mathcal{R}_k + \mathcal{R}_{\neg k}}$
\begin{equation}
w_{i,k} = \frac{\mathcal{R}_{i, \neg k}}{\mathcal{R}_{i, k} + \mathcal{R}_{i, \neg k}}.
\label{eq:weight}
\end{equation}

When $\mathcal{R}_{i,k}+\mathcal{R}_{i,\neg k}=0$ (i.e., all candidates fail across all target languages for $x_i$), the weight is undefined, and in this case, we skip $x_i$ in the policy update.
This weighting scheme dynamically prioritizes lagging directions: if the model demonstrates semantic understanding via sibling languages (high $\mathcal{R}_{i,\neg k}$) but struggles with $L_k$ (low $\mathcal{R}_{i,k}$), $w_{i,k}$ increases, forcing the model to focus on the syntactic or idiomatic hurdles of $L_k$.
We employ the Group Relative Policy Optimization (GRPO) algorithm to optimize the policy with language-aware weighting. The policy is updated by maximizing:
\begin{equation}
\resizebox{\columnwidth}{!}{$
\begin{aligned}
\mathcal{J}(\theta)
= \mathbb{E}\Big[
\sum_{i,k} w_{i,k} \frac{1}{G} \sum_{j=1}^{G} \frac{1}{|o_j|} & \sum_{t=1}^{|o_j|}
\big(
\min(r_{j,t}\hat{A}_{i,k,t}^j, \tilde{c}_{j,t}\hat{A}_{i,k,t}^j)
\\
& - \beta \mathbb{D}_{\mathrm{KL}}(\pi_\theta \,\|\, \pi_{\mathrm{ref}})
\big)
\Big],
\label{eq:grpo-loss}
\end{aligned}
$}
\end{equation}
where $r_{j, t} = \frac{\pi_\theta(y_{i,k,t}^j \,|\, x_{i},y_{i,< t}^{j}, L_k)}{\pi_{\theta_{\text{old}}}(y_{i,k,t}^j \,|\, x_i, y_{i,< t}^{j}, L_k)}$ denotes the importance sampling ratio for the $t$-th token of the $j$-th candidate translation $y_{i,k}^j \in L_k$, and $\tilde{c}_{j,t} = \mathrm{clip}\big(r_{j,t}, 1-\epsilon, 1+\epsilon\big)$
denotes the clipped probability ratio, where $\epsilon$ controls the clipping range.
$w_{i,k}$ is the language-aware weight from Eq.~\eqref{eq:weight}, $|o_j|$ is the length of $y_{i,k}^j$, and $\beta$ is the KL penalty coefficient. 
$\hat{A}_{i,k,t}^j$ is the advantage of $t$-th token in $y_{i,k}^j$ estimated by the same target language group:
\begin{equation}
\resizebox{.89\columnwidth}{!}{$
\hat{A}_{i,k,t}^j = \frac{R\big(y
_{i,k}^j, T\big) - \text{mean}\big(\{ R(y, T) \}_{y \in \mathcal{Y}_{i,k}}\big)}{\text{std}\big(\{ R(y, T) \}_{y \in \mathcal{Y}_{i,k}}\big)}.
$}
\end{equation}

%====================%
\section{Experiments and Results}

\subsection{Experiment Settings}

\textbf{Training data.} The training dataset for \methodname is constructed based on KodCode \cite{zhangchen2025kodcode}. 
We pick the KodCode-RL-10K subset and extract Python solutions and test cases. 
We leverage the MultiPL-E \cite{cassano2022multipl-e} translator to extend these test cases to Java and C++. 
Following its usage, we apply the translation templates to map Python unit-test scaffolds (e.g., entrypoint signatures and assertions) into target-language test harnesses, and discard those translated tests that fail to compile/execute or have ambiguous entrypoint signatures. 
To prevent data leakage, we further remove any training instances whose function names overlap with HumanEval-X or TransCoder-Test.
Ultimately, we curate a dataset comprising 5,584 Python source code, each accompanied by test suites in Python (avg. 8.11 cases), Java (avg. 8.09), and C++ (avg. 8.09).
See Appendix~\ref{app:train} for more details.

\textbf{Implementation.} The training process uses AdamW optimizer with a learning rate of $1\times10^{-6}$. For GRPO, the rollout macro batch size is set to 256 with $G = 8$, and the micro batch size for actor training is 8. The KL penalty coefficient $\beta=0.01$ and the clipping range $\epsilon=0.2$.
During inference, we use greedy decoding for all evaluations to ensure deterministic and reproducible results.

\begin{table*}
    \centering
    \resizebox{\textwidth}{!}{
    \begin{tabular}{l|cccccc|c|cccccc|c}
        \toprule
        \multirow{2}{*}{Methods} & \multicolumn{7}{c|}{HumanEval-X} & \multicolumn{7}{c}{TransCoder-Test} \\ 
        \cmidrule(lr){2-8} \cmidrule(lr){9-15} & P$\to$J & P$\to$C & J$\to$P & J$\to$C & C$\to$J & C$\to$P & Avg & P$\to$J & P$\to$C & J$\to$P & J$\to$C & C$\to$J & C$\to$P & Avg\\
        \midrule
        EffiReasonTrans & 57.32 & 38.41 & 82.32 & 60.37 & \textbf{77.44} & 75.61 & 65.25 & 78.22 & 73.88 & 83.40 & 81.16 & 84.44 & 81.03 & 80.36 \\
		CoTran   & 54.88 & 43.29 & 82.32 & 56.71 & 74.39 & 72.56 & 64.03 & 75.10 & 76.87 & 79.96 & 79.66 & 83.40 & 79.96 & 79.16 \\
        MultiPL-T & 64.02 & 45.73 & 73.78 & 62.20 & 75.00 & 67.68 & 64.74 & 74.48 & \textbf{86.08} & 74.48 & 79.66 & 80.91 & 73.49 & 78.18 \\
        PPOCoder & 68.29 & 54.27 & 82.32 & 62.80 & 73.78 & 73.78 & 69.21 & 75.10 & 79.44 & 80.60 & 86.51 & 84.44 & 80.17 & 81.04 \\
        OORL & 71.34 & 59.15 & 79.27 & 60.98 & 76.83 & 71.95 & 69.92 & 73.86 & 76.02 & 72.84 & 84.37 & 81.74 & 62.50 & 75.22 \\
        \methodname (ours) & \textbf{73.78} & \textbf{60.37} & \textbf{87.20} & \textbf{70.73} & \textbf{77.44} & \textbf{78.66} & \textbf{74.70} & \textbf{79.88} & 85.87 & \textbf{83.62} & \textbf{91.86} & \textbf{84.85} & \textbf{82.11} & \textbf{84.70} \\
        \bottomrule
    \end{tabular}}
    \caption{CA@1 scores of different methods on HumanEval-X and TransCoder-Test, using Qwen3-1.7B.}
    \label{tab:method_main}
\end{table*}

\textbf{Baselines.} We compare our method against two categories of baselines.
(i) \textit{Competing open-source LLMs}. 
We select three widely adopted instruction-tuned checkpoints: Qwen3-32B, Qwen2.5-32B-Instruct, and Llama-3.1-70B-Instruct. They span from 32B to 70B parameters, providing competitive zero-shot code translation capabilities.
(ii) \textit{Representative finetuning methods}:
\begin{itemize}
    \item \textbf{EffiReasonTrans} \cite{wang2025effireasontrans}, a reasoning-enhanced code translation method that conducts RL to optimize CoT paths.
    \item \textbf{CoTran} \cite{jana2024cotran}, a collaborative RL approach that aligns source and target languages by maximizing execution-based rewards and compiler rewards.
    \item \textbf{MultiPL-T} \cite{cassano2023multiplt}, a multilingual code data synthesis framework with a powerful teacher model to generate candidate programs and rejection sampling to curate a dataset for supervised finetuning.
    We synthesize Java and C++ implementations by ensembling Qwen3-32B and Llama-3.1-70B-Instruct, retaining only those passing all test cases. Together with the original Python solutions, this process yields 28,570 translation pairs in total for supervised finetuning.
    \item \textbf{PPOCoder} \cite{shojaee2023executionbased}, an RL-based approach leveraging the PPO algorithm to optimize code translation performance.
    \item \textbf{OORL} \cite{wu2025policy}, a method integrating online RL objectives with offline group DPO training objectives derived from intermediate representations. 
\end{itemize}

These baselines allow us to assess whether \methodname can achieve superior cross-lingual generalization compared to both massive-scale general models and specialized finetuning methods. 
More details are provided in Appendix~\ref{app:impl}.

\begin{table*}
    \centering
    \resizebox{\textwidth}{!}{
    \begin{tabular}{l|cccccc|c|cccccc|c}
        \toprule
        \multirow{2}{*}{Methods} & \multicolumn{7}{c|}{HumanEval-X} & \multicolumn{7}{c}{TransCoder-Test} \\ 
        \cmidrule(lr){2-8} \cmidrule(lr){9-15} & P$\to$J & P$\to$C & J$\to$P & J$\to$C & C$\to$J & C$\to$P & Avg & P$\to$J & P$\to$C & J$\to$P & J$\to$C & C$\to$J & C$\to$P & Avg\\
        \midrule
        \methodname & \textbf{73.78} & \textbf{60.37} & \textbf{87.20} & \textbf{70.73} & \textbf{77.44} & \textbf{78.66} & \textbf{74.70} & \textbf{79.88} & \textbf{85.87} & \textbf{83.62} & \textbf{91.86} & \textbf{84.85} & \textbf{82.11} & \textbf{84.70}\\
		\quad -- Exploration & 70.73 & 56.10 & 81.71 & 62.20 & 76.22 & 77.44 & 70.73 & 79.25 & 83.08 & 81.03 & 88.87 &  84.23 & 80.82 & 82.88 \\
        \quad -- Weighting & 68.90 & 57.93 & 85.98 & 68.29 & 76.22 & 75.61 & 72.16 & 79.57 & 82.66 &  81.90 & 89.94  & 83.61 & 81.47 & 83.19\\
        \bottomrule
    \end{tabular}}
    \caption{CA@1 scores of ablation study with Qwen3-1.7B.}
    \label{tab:ablation}
\end{table*}

\textbf{Evaluation benchmarks and metrics.} We employ the HumanEval-X and TransCoder-Test benchmarks and choose Python, Java, and C++ programming languages. 
See Appendix~\ref{app:bench} for more details.
We use top-1 Computational Accuracy (CA@1) as the efficacy metric.

\subsection{Main Results}
To investigate whether our \methodname can improve code translation accuracy, we compare its performance with its base model on six translation tasks. 

Table~\ref{tab:model_main} presents the experimental results. 
Across all benchmarks and translation tasks, \methodname consistently outperforms its base model. 
% Starting from Qwen3-1.7B, \methodname achieves an average improvement of 10.37\% on HumanEval-X and 5.12\% on TransCode-Test. 
These gains are not confined to translation tasks from Python, showing that \methodname successfully propagates the learning signal beyond the pivot-language data.
A closer look reveals that the weaker the base model, the greater the lift. 
This is precisely the expected pattern of \methodname: 
multilanguage exploration first mines high-confidence translations from the previous iteration and re-feeds them as synthetic source code, instantly expanding the search space; 
subsequent weight optimization then amplifies the learning signal for the weaker translation tasks.

The gains also generalize across different model families. 
As shown in Table~\ref{tab:model_main}, on HumanEval-X and TransCoder-Test, \methodname yields average improvements of 10.37\% and 5.12\% for Qwen3-1.7B, 16.57\% and 5.84\% for Llama-3.1-8B-Instruct, 15.35\% and 1.03\% for Qwen2.5-7B-Instruct, respectively.
Compared with the larger-scale sibling in the same model family, it achieves comparable overall performance and even excels in several specific directions despite much smaller parameters. 

Moreover, we compare \methodname with strong finetuning code-translation baselines in Table~\ref{tab:method_main}. To guarantee a strictly fair comparison, all methods are initialized from the same Qwen3-1.7B base model and trained on the same dataset with \methodname, except for EffiReasonTrans, which uses its released dataset.
\methodname outperforms these methods on both benchmarks on average, with particularly large gains in J$\to$C. 
Overall, these results demonstrate that \methodname is an effective RLVR method for multilingual code translation, leveraging the scalability of unit tests as verifiable oracles.

\subsection{Ablation Study}

%To investigate the effectiveness of each key component in our proposed method, we conduct an ablation study by systematically removing key components and analyzing their impact. 

To measure the individual contributions of bootstrapping multilanguage exploration and language-aware weight optimization, we design two variants:
\begin{itemize}
    \item \textbf{w/o Exploration} removes $\mathcal{D}_{\text{explore}}$, restricting the RL training on the initial pivot seed $\mathcal{D}_\text{seed}$, which only covers P$\to$J/C training data.

    \item \textbf{w/o Weighting} removes the language-aware weighting by setting uniform weights (i.e., $w_{i,k}=1$ for all tasks), and thus all samples contribute equally to the RL objective.
\end{itemize}

Table~\ref{tab:ablation} reports CA@1 on HumanEval-X and TransCoder-Test with Qwen3-1.7B. Removing either component consistently degrades performance across all six directions. 
On HumanEval-X, disabling exploration leads to an average drop of 4\%, showing that bootstrapped multilingual instances are crucial for improving overall performance.
Without the exploration pool, the model can only learn from Python-to-X translations, missing critical reverse and cross-lingual patterns (e.g., J$\to$P, J$\to$C).
This validates that multilingual bootstrapping is essential in evolving beyond pivot-to-X constraints toward broader multilingual translation tasks.

Removing language-aware weighting also decreases performance by 2.5\% on average, indicating that reweighting is crucial for improving hard directions. This effect is most evident on the challenging P$\to$C tasks.
Without adaptive weighting, the model tends to over-optimize on easier translation pairs while neglecting harder ones, leading to imbalanced multilingual proficiency.
The weighting mechanism acts as a curriculum that prevents the policy from getting trapped in exploiting simpler patterns for higher rewards while neglecting the harder tasks.

\begin{table*}
    \centering
    \resizebox{\textwidth}{!}{
    \begin{tabular}{l|cccccc|c|cccccc|c}
        \toprule
        \multirow{2}{*}{Methods} & \multicolumn{7}{c|}{HumanEval-X} & \multicolumn{7}{c}{TransCoder-Test} \\ 
        \cmidrule(lr){2-8} \cmidrule(lr){9-15} & P$\to$J & P$\to$C & J$\to$P & J$\to$C & C$\to$J & C$\to$P & Avg & P$\to$J & P$\to$C & J$\to$P & J$\to$C & C$\to$J & C$\to$P & Avg\\
        \midrule
        \methodname & 73.78 & 60.37 & 87.20 & 70.73 & 77.44 & 78.66 & 74.70 & 79.88 & 85.87 & 83.62 & 91.86 & 84.85 & 82.11 & 84.70 \\
        \quad + InterTrans & 84.15 & 76.22 & 90.85 & 76.22 & 84.15 & 84.76 & 82.73 & 86.31 & 88.65 & 84.91 & 93.15 & 88.80 & 83.19 & 87.50 \\
		\quad + UniTrans & 78.05 & 60.98 & 87.80 & 71.34 & 80.49 & 81.71 & 76.73 &  89.83 & 88.87 & 87.72 & 93.36 & 91.70 & 86.85 & 89.72 \\
        \bottomrule
    \end{tabular}}
    \caption{Compatibility with inference-time enhancement strategies.}
    \label{tab:compatibility}
\end{table*}

\subsection{Different Pivot Languages}

To see how the choice of pivot language influences performance, we repeat the full training pipeline of \methodname with Qwen3-1.7B by substituting the default Python pivot with Java and C++.
Since the original training split contains only Python code as the starting point, we adopt the synthesized Java and C++ code used in MultiPL-T.
This includes 5,254 Java and 4,555 C++ source programs, which serve as cross-lingual counterparts to a subset of the original Python references for training.
We categorize these as ``silver-standard'' references compared to the ``gold-standard'' Python seeds.

\begin{figure}
    \centering
    \includegraphics[width=\linewidth]{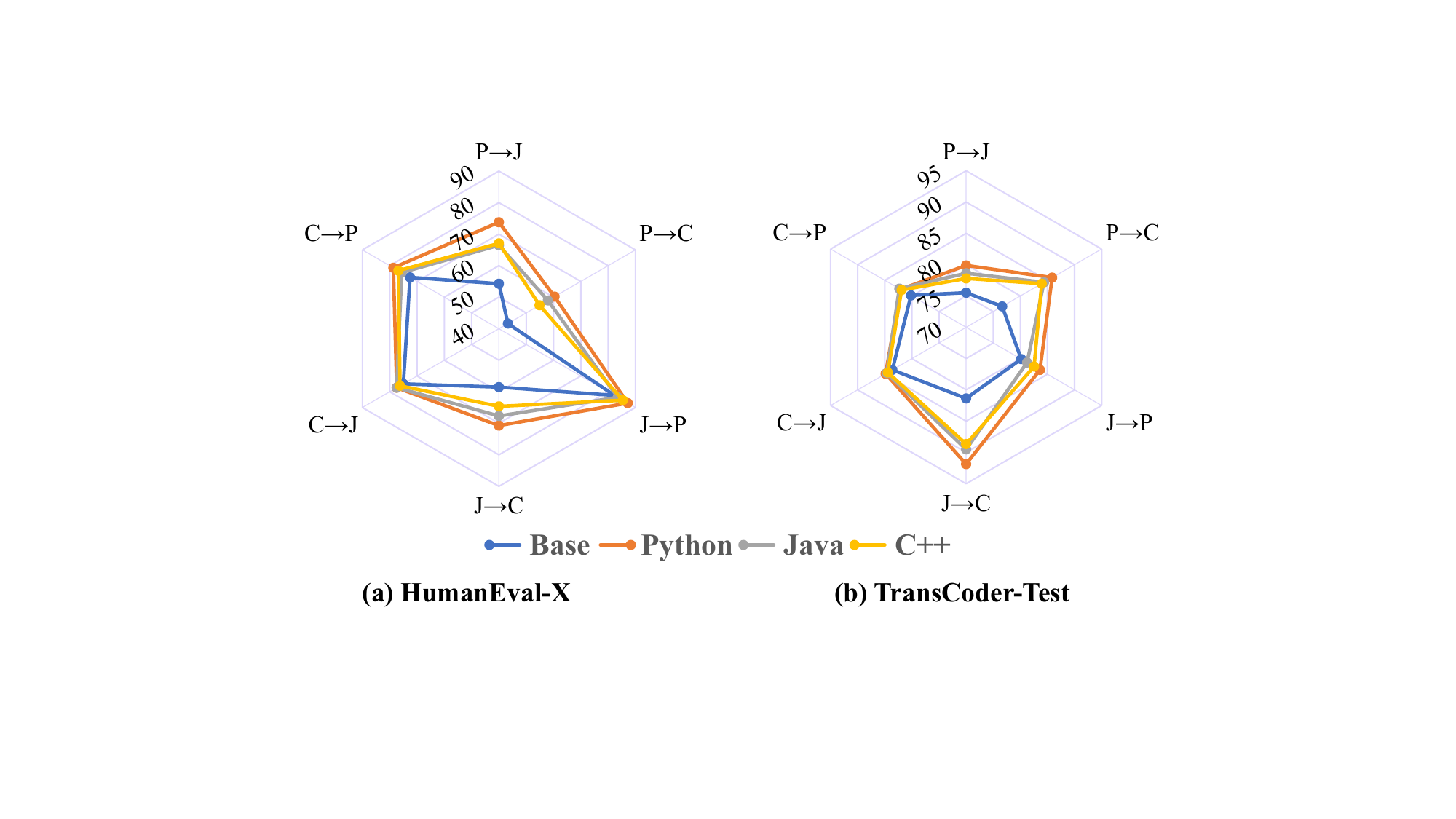}
    \caption{Performance comparison of different pivot languages with Qwen3-1.7B.}
    \label{fig:pivot}
\end{figure}

As depicted in Figure~\ref{fig:pivot}, our method achieves consistent performance gains compared with the base model regardless of the chosen pivot language, indicating that the core mechanism of \methodname is effective.
The radar chart visualizes the CA@1 scores across all six translation directions on both HumanEval-X and TransCoder-Test benchmarks, where each axis represents a specific translation task.
While \methodname is language-agnostic, using Python as the pivot yields the superior overall performance, with the largest coverage area in the radar chart.
We attribute this advantage to two factors. 
The first advantage stems from the fact that our Python seed data is inherently more abundant and covers a significantly broader range of algorithmic logic and functional scenarios, providing a higher-quality initialization for exploration.
Second, LLMs are typically optimized with a higher proportion of Python corpora during pre-training. 
Consequently, starting the exploration from Python leverages the model's strongest internal representations, providing more potential for cross-lingual knowledge transfer.

\subsection{Compatibility to Existing Framework}

Recent breakthroughs in code translation focus on inference-time strategies, which are orthogonal to the training-time optimizations of \methodname.
To further demonstrate the compatibility of \methodname to inference-time enhancement strategies, we evaluate it by integrating it with two distinct paradigms: 
\begin{itemize}
\item Path-based scaling via \textbf{InterTrans} \cite{Macedo2025InterTrans}, which explores multiple translation routes. 
This method capitalizes on the model's inherent multilingual translation proficiency by expanding the sampling budget.
The performance of InterTrans serves as a direct indicator of the base model's multilingual code translation effectiveness.

\item Iterative refinement via \textbf{UniTrans} \cite{yang2024exploring}, which utilizes execution feedback to progressively refine the output.
It exploits the model's self-refine ability.
\end{itemize}

\begin{table}
\centering
\resizebox{\columnwidth}{!}{
\begin{tabular}{l|cccccc}
\toprule
Unseen & J$\rightarrow$Go & C$\rightarrow$Go & JS$\rightarrow$Go & Go$\rightarrow$J & Go$\rightarrow$C & Go$\rightarrow$JS \\
\midrule
Qwen3-1.7B & 45.73 & 54.88 & 53.05 & 70.12 & 52.44 & 92.07 \\
\ \ +\methodname & \textbf{60.37} & \textbf{57.93} & \textbf{66.46} & \textbf{75.0} & \textbf{57.93} & \textbf{96.34} \\
\midrule
Low-res. & P$\rightarrow$D & P$\rightarrow$R & J$\rightarrow$D & J$\rightarrow$R & C$\rightarrow$D & C$\rightarrow$R \\
\midrule
Qwen3-1.7B & 23.08 & 12.18 & 21.79 & 13.46 & 25.00 & 16.03 \\
\ \ +\methodname & \textbf{41.67} & \textbf{28.85} & \textbf{33.33} & \textbf{23.08} & \textbf{41.67} & \textbf{25.64} \\
\bottomrule
\end{tabular}
}
\caption{CA@1 scores of extended language pairs. ``JS'', ``D'', and ``R'' denote JavaScript, Dlang, and Racket, respectively.}
\label{tab:morelang_performance}
\end{table}

\begin{table*}
\centering
\resizebox{\textwidth}{!}{
\begin{tabular}{l|cc|cc|cc|cc|cc|cc}
\toprule
\multirow{2}{*}{Models} & \multicolumn{2}{c|}{C$\rightarrow$P} & \multicolumn{2}{c|}{J$\rightarrow$P} & \multicolumn{2}{c|}{P$\rightarrow$J} & \multicolumn{2}{c|}{C$\rightarrow$J} & \multicolumn{2}{c|}{P$\rightarrow$C} & \multicolumn{2}{c}{J$\rightarrow$C} \\
\cmidrule(lr){2-3} \cmidrule(lr){4-5} \cmidrule(lr){6-7} \cmidrule(lr){8-9} \cmidrule(lr){10-11} \cmidrule(lr){12-13}
 & $CA_c$ & $CA_m$ & $CA_c$ & $CA_m$ & $CA_c$ & $CA_m$ & $CA_c$ & $CA_m$ & $CA_c$ & $CA_m$ & $CA_c$ & $CA_m$ \\ \midrule
Qwen3-1.7B & 18.09 & 28.72 & 17.02 & 25.53 & 3.19 & 30.85 & 6.38 & 38.30 & 4.26 & 10.64 & 0.00 & 0.00 \\
\methodname Qwen3-1.7B & \textbf{23.40} & \textbf{35.11} & \textbf{19.15} & \textbf{32.98} & \textbf{8.51} & \textbf{46.81} & \textbf{9.57} & \textbf{56.38} & \textbf{4.26} & \textbf{10.64} & \textbf{0.00} & \textbf{1.06} \\ \midrule
Qwen2.5-7B-Instruct & 36.17 & 48.93 & 28.72 & \textbf{38.30} & 4.26 & 32.98 & 4.26 & 32.99 & 3.19 & 6.38 & \textbf{1.06} &  \textbf{1.06}\\
\methodname Qwen2.5-7B-Instruct & \textbf{40.43} & \textbf{52.13} & \textbf{29.79} & \textbf{38.30} & \textbf{18.08} & \textbf{68.09} & \textbf{14.89} & \textbf{74.47} & \textbf{4.26} & \textbf{9.57} & \textbf{1.06} & \textbf{1.06}\\ \bottomrule
\end{tabular}
}
\caption{$CA_c$ and $CA_m$ scores on ClassEval-T benchmark.}
\label{tab:classevalt}
\end{table*}

As presented in Table~\ref{tab:compatibility}, \methodname serves as a solid foundation for InterTrans and UniTrans. 
While \methodname improves the success rate at first attempt, the integration with InterTrans and UniTrans further unlocks its potential.
Specifically, when integrated with InterTrans, the CA@1 scores exhibit a substantial further improvement, providing more reliable candidates across diverse translation paths.

Furthermore, the integration with UniTrans reveals that despite undergoing a translation-specific RL process, the model preserves its intrinsic self-refine ability.
This suggests that \methodname is additive to existing test-time compute-scaling and feedback-driven methods.

\subsection{Language Extension}
To further validate the generalization of \methodname, we extend our evaluation to a broader spectrum of programming languages on the HumanEval-X benchmark. 
% We conduct all experiments using the Qwen3-1.7B model on the HumanEval-X benchmark.
Specifically, we investigate whether \methodname can effectively generalize to unseen languages, which are not encountered during the training phase. 
We select Go as a representative. 
Furthermore, acknowledging that LLMs often falter when dealing with low-resource languages due to the scarcity of training corpora \cite{rosel2024finetunelow}, we assess the performance of \methodname on Dlang and Racket. 
For these low-resource scenarios, we leverage the MultiPL-E \cite{cassano2022multipl-e} to extend the test cases.

As illustrated in Table \ref{tab:morelang_performance}, \methodname consistently outperforms the base Qwen3-1.7B model across all tested pairs. 
Notably, in the low-resource scenarios such as P$\to$D, \methodname achieves a significant performance uplift, demonstrating its superior capability in bridging the gap for languages with limited data availability. 
% This performance leap suggests that \methodname generalizes effectively, even when the language's footprint is minimal.
This performance leap suggests that \methodname effectively enhances cross-lingual alignment capabilities, even for languages that are not explicitly optimized during training or those with a minimal data footprint.

\subsection{Class-level Code Translation}
To further evaluate the robustness and generalizability of \methodname in complex, real-world scenarios, we conduct an additional experiment on ClassEval-T \cite{xue2025classeval}, a significantly more challenging benchmark for class-level code translation.
Unlike conventional method-level or program-level tasks, the benchmark involves longer contexts, intricate cross-method dependencies, and external library calls, requiring the model to maintain consistency across an entire class.
Adopting the standard protocol for this benchmark, we employ class-level CA ($CA_c$) and method-level CA ($CA_m$) to evaluate performance. 

As shown in Table~\ref{tab:classevalt}, even without direct training on class-level data, \methodname achieves improvements across most translation pairs. 
We observe no gains in the J$\to$C task, which we attribute to the inherent capacity limits of the base model in handling C++'s complex memory management and syntax at scale. 
% In future, we plan to explore agentic RL methods to better handle such "knowledge void" issue.

\begin{figure*}
    \centering
    \includegraphics[width=\linewidth]{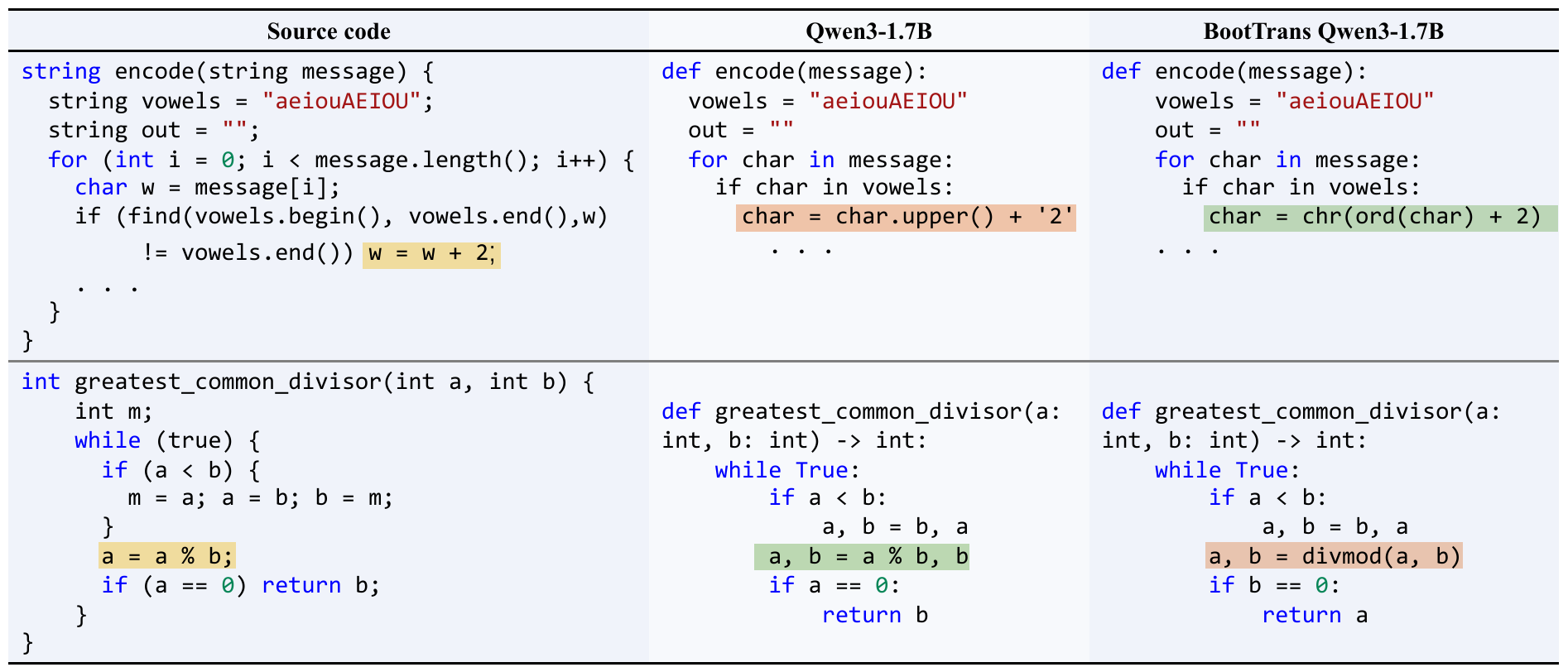}
    \caption{Two examples of C++ translated into Python. \methodname succeeds in the top case but fails in the bottom.}
    \label{fig:case}
\end{figure*}

\subsection{Case Study}
To investigate the qualitative impact of \methodname beyond numerical metrics, we analyze specific translation instances to understand how \methodname shifts the model's translation behavior.

As shown in Figure~\ref{fig:case}, in the top string encoding case, the C++ expression \textit{w + 2} increments the ASCII code of w by 2, yielding another single character. However, Qwen3-1.7B translates it into \textit{+ `2'}, which is a string concatenation operation. It alters both the length and the content of the output.
In contrast, \methodname faithfully reproduces the original source C++ code's behavior with \textit{ord/chr + 2 }. 
This demonstrates that \methodname preserves the exact logic of the source code.

The bottom GCD case highlights the aggressive nature of our exploration strategy. 
Qwen3-1.7B remains a literal translation of \% operator.
Conversely, \methodname uses the Pythonic \textit{divmod} idiom. 
Despite misalignment in return types, \methodname empowers the model to search for higher-level algorithmic equivalents.
% and improved computational efficiency.
Collectively, both cases suggest that \methodname generates more idiomatic and native-like translations. 
While exploration entails risks, as seen in the idiomatic misuse, it represents a vital opportunity for achieving more complex translations, such as built-in function mapping and API adaptation.
See Appendix~\ref{app:error} for more details about the categorization of translation errors.

%====================%
\section{Related Work}

Code translation research has evolved through multiple paradigms. Early unsupervised methods, e.g., TransCoder \cite{roziere2020transcoder}, TransCoder-ST \cite{roziere2022leveraging}, and structure-aware variants \cite{szafraniec2022code,huang2023program,liu2023syntax} avoid parallel data dependency but require massive corpora and heavy computation. 
With the availability of larger parallel corpora \cite{zhu2022xlcost,yan2023codetransocean,mohammad2024xcodeeval,weixiang2024codescope} and pre-trained models \cite{wang2021codet5,guo2021graphcodebert,zheng2023codegeex,feng2020codebert}, supervised finetuning has become the dominant paradigm.
Recent work emphasizes executable, semantics-oriented evaluation: MultiPL-E \cite{cassano2022multipl-e} enables scalable compile-and-run testing; reliability analyses and richer testing further study functional correctness \cite{pan2024lost,eniser2024testing}. 

Beyond supervised finetuning, RLVR offers a promising direction. 
CodeRL \cite{le2022coderl} pioneers execution-based signals for code generation, while CoTran \cite{jana2024cotran}, PPOCoder \cite{shojaee2023executionbased}, and OORL \cite{wu2025policy} integrate compiler and symbolic-execution feedback. 
EffiReasonTrans \cite{wang2025effireasontrans} combines reasoning augmentation with RL to balance accuracy and latency. These methods apply PPO or GRPO \cite{schulman2017proximal,shao2024deepseekmath} for reward-based optimization. 
However, existing RLVR methods often rely on uniform optimization across language pairs and may be sensitive to test coverage.
\methodname addresses the challenges of data scarcity and multilingual optimization imbalance through bootstrapping multilanguage exploration and language-aware weighting.
Test-time improvements such as explanations, iterative feedback or transitive intermediate translations can also boost quality \cite{tang2023explainthen,Macedo2025InterTrans, yang2024exploring}, motivated by multilingual software co-evolution \cite{zhang2023coevolution}.

%====================%
\section{Conclusion} 

This paper proposes \methodname, a novel method for multilingual code translation to address data scarcity and optimization imbalance challenges.
By leveraging functional invariance and portable test oracles, we establish universal oracles for multilingual RL training. 
We integrate a dual-pool architecture and a language-aware weighting mechanism to bootstrap multilanguage exploration and achieve a balanced optimization.
Our experiments demonstrate substantial improvements over strong baselines, with ablations confirming the effectiveness of both key components.
Future work will extend \methodname to functional languages and explore more sophisticated reward mechanisms.

\section*{Acknowledgments}
This work was supported by the National Natural Science Foundation of China (No. 62272219), the Fundamental and Interdisciplinary Disciplines Breakthrough Plan of the Ministry of Education of China (No. JYB2025XDXM118), and the Cooperation Fund of Huawei Cooperation Project (No. TC20230202021-2024-12).

%====================%
% 不占页数限制
\section*{Ethical Considerations}
The datasets, benchmarks, and LLMs used in this paper are public with permissive licenses.

\section*{Limitations}
While \methodname demonstrates substantial improvements in multilingual code translation, several limitations warrant discussion.
First, our evaluation mainly focuses on three imperative languages (Python, Java, C++) and may require additional adaptation for domain-specific languages with fundamentally different paradigms.
% Second, the quality of bootstrapped translations depends on the comprehensiveness of unit tests; incomplete test coverage may allow functionally incorrect translations to enter the exploration pool, potentially propagating errors.
Second, our language-aware weighting relies on execution-based binary rewards, which may not capture nuanced aspects of code quality such as readability, maintainability, or idiomatic style.
% Finally, the choice of pivot language significantly impacts overall performance, and our empirical results suggest Python's dominance may not generalize to scenarios where other languages have richer initial resources or stronger model priors.
Finally, the efficacy of \methodname is partially influenced by the test suite size as shown in Appendix~\ref{app:sense}.
A scarcity of test cases in the initial pivot language may lead to a performance degradation.

% Bibliography entries for the entire Anthology, followed by custom entries
%\bibliography{anthology,custom}
% Custom bibliography entries only
\bibliography{custom}

%====================%
\appendix

\section{Details of Training Data Construction}
\label{app:train}
In the original KodCode dataset, not all Python code snippets are accompanied by function signatures. We use Qwen3-32B to annotate the function signatures, verify the correctness of these annotated function signatures, and ultimately retain only the Python source code snippets with valid function signatures.
The test cases provided with the Python code in the original KodCode dataset have all been verified to be correct. Therefore, we directly use MultiPL-E to translate these test cases, and filter out the test cases that failed to be translated (the proportion of such failed cases is very small, only 0.2\%).
To avoid data leakage, we filtered out all Python code snippets whose function entry points overlap with those in HumanEval-X and TransCoder-Test. Finally, we obtained a total of 5,584 valid data samples.

\section{Details of Baseline Implementation}
\label{app:impl}
Unlike prior studies that assume the availability of parallel training data, our setting operates under a monolingual-pivot regime. To ensure a fair comparison, we adapt these baselines to align with our non-parallel training constraints while preserving their original algorithmic essence.

For EffiReasonTrans, it relies on costly chain-of-thought distillation from DeepSeek-R1 through multi-step reasoning annotations. 
Due to the prohibitive expense of reproducing this pipeline, we instead utilize its publicly released reasoning-augmented dataset to ensure its competitive edge. We first perform supervised finetuning, followed by RL using GRPO, maintaining the same reward function and hyperparameter configurations as \methodname for parity.

For MultiPL-T, which originally uses a single teacher model (e.g., StarCoder) for data synthesis, we enhance its data curation process by employing two powerful open-source models, Llama-3.1-70B and Qwen3-32B, to perform rejection sampling on \methodname's seed dataset. 
Specifically, we generate candidate translations from both models for the to-Java and to-C++ tasks and retain only valid translations that pass all unit tests. 
We then pair these translations with Python in seed dataset and construct a clean parallel code translation dataset for superised finetuning. This modified pipeline better aligns with our non-parallel training constraint while preserving the core idea of MultiPL-T.

For CoTran, we follow the original implementation, including its forward–backward policy architecture and reward formulation. However, we observe a significant performance degradation compared to both our method and other adapted baselines. We attribute this to a training-reward mismatch: the backward policy is optimized using the forward policy's outputs without execution-based filtering. Consequently, the resulting reward signal becomes noisy and misleading, as the backward policy may reward translations that are reconstructible but functionally incorrect.

For PPOCoder, its original reward mechanism relies on reference translations, which are unavailable in our setting. To adapt it fairly, we replace its reference-based reward with our unit test pass rate (execution feedback). 
We then apply the standard PPO algorithm to optimize the base model, serving as a representative baseline for conventional RL-based code translation without parallel supervision.

For OORL, we use the same dataset as \methodname for the online RL component. For the offline component, since the original data is not publicly available, we randomly select 9,600 curated C-to-IR (intermediate representation) groups from the SLTrans dataset.

\section{Details of Benchmarks}
\label{app:bench}
We select HumanEval-X and TransCoder-Test as the evaluation datasets, due to their widespread recognition and adoption within the research community.
The HumanEval-X dataset contains 164 source code snippets for each programming language. 
Thus, across the six cross-translation tasks among C++, Python, and Java, it constitutes a total of 984 test samples, with an average of 6.9 test cases per sample.
For the TransCoder-Test dataset, the number of samples for the tasks of translating to Java, Python, and C++ are 482, 464, and 467, respectively. 
It constitutes a total of 2,826 test samples, with an average of 10 test cases per sample.

\section{Details of Prompt}
\label{app:prompt}
We follow the code translation setting of HumanEval-X, where we leverage declarations (shared across both source and target languages) and translate the solution from the source language to the target language. The prompt template adopted for model training and inference is below:

\begin{promptbox}{Prompt}
\begin{small}
\begin{verbatim}
Please translate source {{source_lang}} 
code to target {{target_lang}} code:
```{{source_lang}}
{{source_code}}```
The translated {{target_lang}} code should 
be:
```{{target_lang}}
{{target_signature}}```
\end{verbatim}
\end{small}
\end{promptbox}

\section{Classification of Translation Errors}
\label{app:error} 
To analyze how \methodname alters the model’s translation behavior, we categorize translation errors into four primary types:
\begin{itemize}
\item \textbf{Logical inconsistency:} The translated code exhibits different runtime behavior or logic compared to the source code;

\item \textbf{Syntactic invalidity:} The generated code violates the grammatical rules of the target programming language and fails to compile or parse;

\item \textbf{API misuse:} Inappropriate usage of APIs in the target language, such as using deprecated functions, wrong argument order, or non-existent library calls;

\item \textbf{Type mismatch:} The translated code employs data types or structures that are incompatible with the target language's type system or the intended operations, such as improper casting or incorrect collection types;

\item \textbf{Signature mismatch:} Discrepancies between the generated function signatures and the expected entry point in the unit tests,  which prevent the code from being correctly invoked.
\end{itemize}

\begin{figure}
    \centering
    \includegraphics[width=0.95\linewidth]{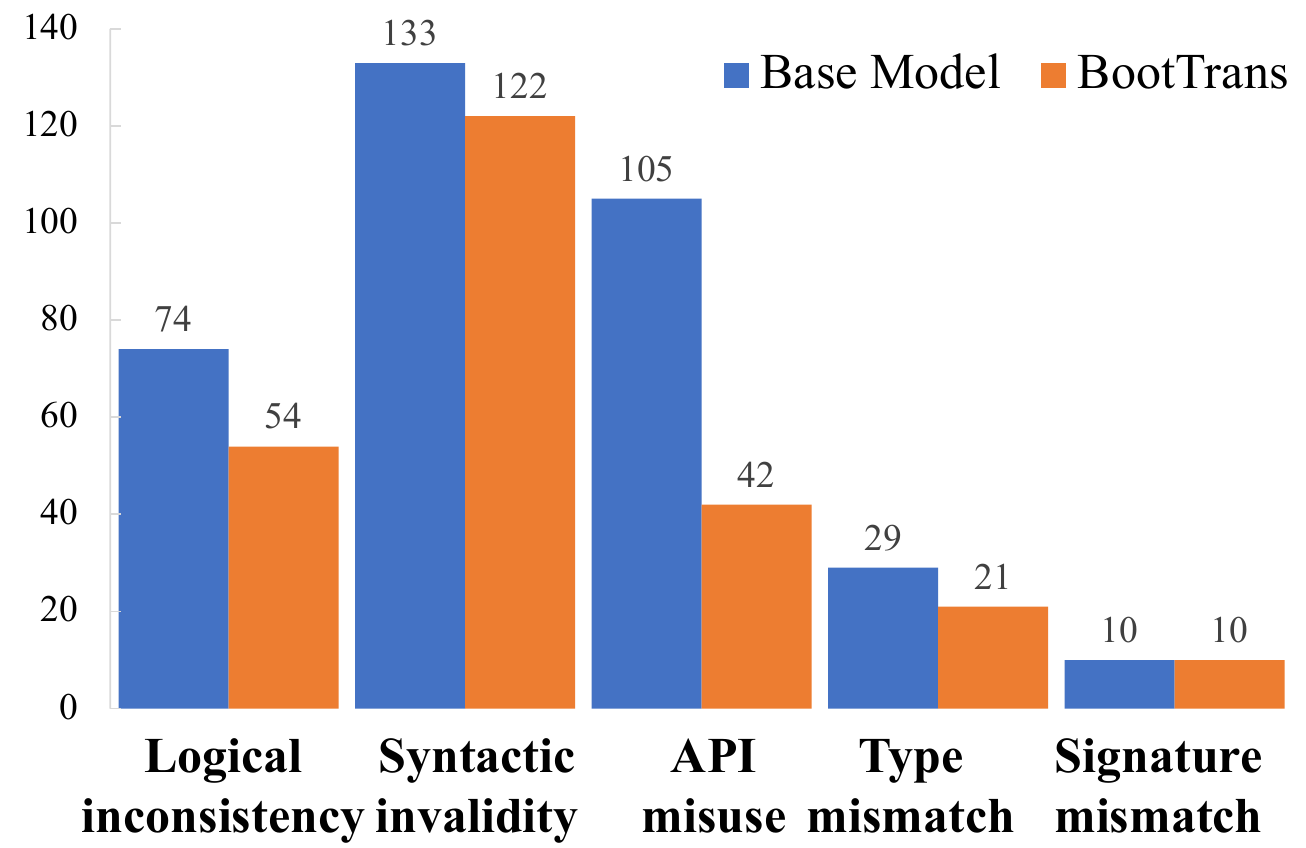}
    \caption{Error classification on HumanEval-X.}
    \label{fig:errors}
\end{figure}

As shown in Figure~\ref{fig:errors}, the error distribution analysis reveals that \methodname significantly outperforms the base Qwen3-1.7B model across multiple failure modes, particularly reducing API misuse and logical inconsistency. The dramatic 60\% reduction in API-related errors underscores the effectiveness of \methodname. 

\section{Sensitivity to Unit Tests Scale}
\label{app:sense} 
Since \methodname relies on execution-based feedback as the primary signal for RL and data bootstrapping, the quality and quantity of test cases are critical. 
To investigate the \methodname's sensitivity to the comprehensiveness of the verification oracle, we conduct experiments using different subsets of the original test suites. 
Specifically, we subsample the test cases for each problem to 50\% and 25\% of their original sizes randomly.

Based on the full set of test cases, \methodname achieves the highest accuracy across all languages, as the dense test suite provides the most rigorous ``grounding'' for functional correctness. 
When the test cases are reduced to 50\%, the average performance remains remarkably stable compared to the full-set setting; however, the impact on individual translation directions is mixed. 
The robustness at 50\% test scale suggests that a representative subset is sufficient to capture the core functional logic. \methodname effectively leverages this high-quality sparse feedback to maintain reliable weighting $w_{i, k}$. 
The 50\% scale acts as a natural regularizer; it smooths the reward landscape and prevents extreme weight polarization. 
This leads to a more balanced performance.
When the test cases are reduced to 25\%, we observe a moderate performance degradation. 

\begin{table}[!ht]
\centering
\resizebox{\linewidth}{!}{
\begin{tabular}{l|cccccc|c}
\toprule
& P$\to$J & P$\to$C & J$\to$P & J$\to$C & C$\to$J & C$\to$P & Avg \\
\midrule
Full & 73.78 & 60.37 & 87.20 & 70.73 & 77.44 & 78.66 & 74.70 \\
50\% & 74.39 & 65.85 & 84.15 & 72.56 & 72.56 & 78.66 & 74.70 \\
25\% & 73.17 & 59.15 & 83.54 & 70.12 & 75.00 & 75.00 & 72.66 \\
\bottomrule
\end{tabular}}
\caption{Sensitivity analysis of \methodname w.r.t.~test suite size on HumanEval-X. We compare the translation performance on HumanEval-X using the full test suite against reduced versions containing 50\% and 25\% of the original test cases.}
\label{tab:test cases}
\end{table}

\end{document}